%% file: main.tex
\pdfoutput=1
\documentclass[sigconf]{acmart}
\usepackage{balance}
\def\BibTeX{{\rm B\kern-.05em{\sc i\kern-.025em b}\kern-.08emT\kern-.1667em\lower.7ex\hbox{E}\kern-.125emX}}

\usepackage{bm}
\usepackage{amssymb}
\usepackage{multirow}
\usepackage{colortbl}
\usepackage{xcolor}
\usepackage{tikz}
\usetikzlibrary{shapes.misc}
\tikzset{
draw/.append style={font=\normalsize},
font={\fontsize{8pt}{12} \fontfamily{libertine} \selectfont},
cross/.style={
cross out, draw, minimum size=2*(#1-\pgflinewidth), inner sep=0pt, outer sep=0pt
}
}
 
\copyrightyear{2019}
\acmYear{2019}
\setcopyright{acmlicensed}
\acmConference[MM '19]{Proceedings of the 27th ACM International Conference on Multimedia}{October 21--25, 2019}{Nice, France}
\acmBooktitle{Proceedings of the 27th ACM International Conference on Multimedia (MM '19), October 21--25, 2019, Nice, France}
\acmPrice{15.00}
\acmDOI{10.1145/3343031.3350873}
\acmISBN{978-1-4503-6889-6/19/10}

\acmSubmissionID{1196}

\newcommand{\colornull}{ACMGreen}
\newcommand{\colorone}{ACMYellow}
\newcommand{\colortwo}{ACMRed}
\newcommand{\colorthree}{ACMDarkBlue}
\newcommand{\colorfour}{ACMLightBlue}
\newcommand{\colorfive}{ACMBlue}
\newcommand{\colorsix}{ACMPurple}
\newcommand{\colorseven}{ACMOrange}
\newcommand{\coloreight}{gray}
\newcommand{\colornine}{pink}

\begin{document}

\title[Audiovisual Transformer Architectures]{Audiovisual Transformer Architectures for \\ Large-Scale Classification and Synchronization of \\ Weakly Labeled Audio Events}

\author{Wim Boes}
\email{wim.boes@esat.kuleuven.be}
\author{Hugo Van hamme}
\email{hugo.vanhamme@esat.kuleuven.be}
\affiliation{
\institution{ESAT, KU Leuven}
\city{Leuven}
\state{Belgium}
}

\begin{abstract}
We tackle the task of environmental event classification by drawing inspiration from the transformer neural network architecture used in machine translation. We modify this attention-based feedforward structure in such a way that allows the resulting model to use audio as well as video to compute sound event predictions. 

We perform extensive experiments with these adapted transformers on an audiovisual data set, obtained by appending relevant visual information to an existing large-scale weakly labeled audio collection. The employed multi-label data contains clip-level annotation indicating the presence or absence of 17 classes of environmental sounds, and does not include temporal information.

We show that the proposed modified transformers strongly improve upon previously introduced models and in fact achieve state-of-the-art results. We also make a compelling case for devoting more attention to research in multimodal audiovisual classification by proving the usefulness of visual information for the task at hand, namely audio event recognition.

In addition, we visualize internal attention patterns of the audiovisual transformers and in doing so demonstrate their potential for performing multimodal synchronization.
\end{abstract}



\maketitle

\section{Introduction}
Human beings heavily depend on both auditory and visual cues to detect environmental events and to infer how to act in response. In order to create situation-aware machines, it is therefore vital to study and design models which are capable of utilizing both modalities to accurately classify such events.

Image classification has been a red hot topic ever since the advent of the ImageNet challenge and data set. Much more recently, research into audio event classification has also taken off because of among other the introduction of the Audio Set data set and challenges such as DCASE 2016, DCASE 2017 and DCASE 2018. However, in contrast to the unimodal classification tasks, multimodal classification tasks are much less prevalent in the literature, even though studying those endeavours could naturally lead to important practical consequences \cite{ImageNet, AudioSet, DCASE2016, DCASE2017proceedings, DCASE2018proceedings}. 

In this work, we draw inspiration from the machine translation community to tackle classification of environmental events. As is the case with many machine learning tasks, the domain of machine translation is currently dominated by the deep learning paradigm. In particular, recently the transformer neural network has gained a lot of popularity. This type of model attains state-of-the-art performance while also achieving comparatively quick training times and allowing for interpretation relatively easily \cite{transformers}.

The data used in the field of machine translation is similar to audiovisual data in the sense that two modalities are involved, namely a source language and a target language. The comparison can be taken even further as audio and vision could be regarded as two abstract languages capable of describing environmental events. For example, a train passing by could be expressed in the abstract audio language by the sound of a train horn, while it could simultaneously be represented in the abstract vision language by a picture of a train. This loose analogy indicates that it could be useful to apply machine translation techniques to audiovisual data.

The approach taken in this work is therefore the following: we modify the transformer neural network architecture used in machine translation to be viable for the task of audiovisual classification of environmental events. We then evaluate and analyze the results obtained by these models.

In Section~\ref{sect:multtrans}, we discuss the original transformer architecture and how it can be adapted for the task of audiovisual classification of environmental events. In Section~\ref{sect:expset}, we discuss the setup for the conducted experiments. Next, in Section~\ref{sect:results} we discuss and interpret the obtained results. Finally, in Section~\ref{sect:concl} we draw a conclusion.

\section{Audiovisual transformer \\  neural networks}
\label{sect:multtrans}

In this section, we discuss some key concepts of the original transformer used for machine translation and explain how this model can be adapted to become suitable for audiovisual classification.

\subsection{Machine translation transformer}

The original transformer is a neural network based entirely on feedforward layers, residual connections and attention. It is depicted in Figure~\ref{fig:mttrans}. This encoder-decoder model converts a sequence of input tokens in a source language into a sentence in a target language by iteratively calculating conditional probabilities of each output word and combining them through beam search. In what follows, we will further describe some of the key components of this model. However, we will not discuss elements which are not used in this work such as the masking and shifting of output tokens and implementation details like the used regularization methods. For full details on the machine translation transformer, we refer the reader to the original work on this model \cite{transformers, residual}. 

\input{mttrans}

\subsubsection{Multi-head attention}

The transformer in Figure~\ref{fig:mttrans} uses multi-head scaled dot product attention blocks. Scaled dot product attention involves computing linear combinations of a set of values. The so-called attention weights of these sums are determined by comparing a number of queries to a collection of keys corresponding to the values: they are obtained by calculating scaled dot products and applying the softmax function to normalize w.r.t. the keys and values. Mathematically, this can be expressed as follows \cite{transformers}: 

\begin{equation}
A(Q, K, V) = \text{softmax}\left(\dfrac{QK^T}{\sqrt{d_k}}\right)V
\label{eq:sdpatt}
\end{equation}

In Equation~\eqref{eq:sdpatt}, $Q \in \mathbb{R}^{n_q \times d_k}$, $K \in \mathbb{R}^{n_k \times d_k}$ and $V \in \mathbb{R}^{n_k \times d_v}$ denote the matrices containing respectively the queries, the keys and the values. The number of queries and the number of keys, which is equal to the number of corresponding values, are symbolized by $n_q$ and $n_k$ respectively. The dimension of the keys, which is the same as the dimension of the queries, and the dimension of the values are indicated by $d_k$ and $d_v$ respectively. The matrix $A \in \mathbb{R}^{n_q \times d_v}$ contains the resulting weighted sums \cite{transformers}.

The multi-head attention blocks add flexibility on top of this bare attention mechanism in two ways. Firstly, linear transformations are introduced before and after the computation of the scaled dot products. Secondly, instead of one linear combination, multiple attention heads, i.e., weighted sums, are calculated and concatenated per query, explaining the name of the component \cite{transformers}.

The multi-head attention block is zoomed in on in Figure~\ref{fig:mhatt}.

\input{mhatt}

In the transformer architecture for machine translation, the queries, keys and values consist of abstract word token representations. It has been demonstrated that applying the multi-head attention mechanism to these word embeddings in both the encoder and decoder allows the model to capture a variety of interesting intra- as well as interlingual word relationships \cite{transformers}.

\subsubsection{Positional encodings} 

Because of the absence of recurrent and convolutional connections, transformers have no real perception of order, which is obviously an undesirable property for neural machine translation models. This issue is alleviated by injecting sequential information via the introduction of positional encodings. This comes down to adding abstract vector representations of word positions to both the input and output word embeddings. 
In the machine translation transformer, sinusoidal positional encodings are employed: for this type of encodings, fixed position differences can be represented by simple linear transformations \cite{posenc, transformers}.

\subsection{Adaptations to original architecture}

Due to the similarities between machine translation and audiovisual classification, fitting the original transformer to the task at hand is relatively easy. Below, we discuss the relevant changes.

The resulting audiovisual transformer architecture incorporating all of these adaptations is depicted in Figure~\ref{fig:avtrans}.

\subsubsection{Data adaptation}

Linguistic data is sequential in nature, defined at discrete word positions. In this work, we ensure that the used audiovisual data complies to this format by converting the audio into sequences of frames and sampling images from the videos. More detailed information is provided in Section~\ref{sect:expset}.

In machine translation, the data from the source language is served as inputs, while the data from the target language constitutes the desired outputs. 
For the task at hand, all data from the considered modalities, i.e., audio and vision, can be used as inputs. 

\subsubsection{Introduction of extra linear maps}

For the machine translation transformer, the size of the word embeddings constrains the amount of neurons used by the linear transformations in the multi-head attention blocks. This is a consequence of the residual connections in the model. To relax this restriction, the audiovisual transformer contains additional linear maps projecting the employed auditory and visual embeddings to vectors of some predefined dimensionality. 

\subsubsection{Change of output layer}

In contrast to machine translation, for classification of environmental events the prediction categories are not mutually exclusive. Indeed, while word positions can only be occupied by a single token, multiple environmental events can occur at the same time. The softmax layer at the end of the transformer is therefore changed into a sigmoid layer as it is more appropriate.

\subsubsection{Addition of aggregation block}

The goal of the machine translation transformer is to compute the probability of a sentence. This requires iteration as conditional probabilities have to be calculated for each word in order to eventually multiply them. However, for the task at hand and the used data (see Section \ref{sect:expset}), we only need a single probability vector per audiovisual clip. Thus, we add an aggregation block at the end of the model which combines frame-level scores by performing a mean or max pooling operation. Consequently, unlike in the original transformer, there is no more need for iteration in
the adapted variant: inputs need to be passed through this model just once to obtain sound event predictions.

\subsubsection{Modification of attention function}

Originally, the softmax function was used to calculate attention values as is evident from Equation~\eqref{eq:sdpatt}. We also test some other straightforward attention functions, namely the sigmoid function $\sigma$ and the normalized sigmoid function $\overline{\sigma}$. The latter function is characterized as follows:

\begin{equation}
\overline{\sigma}(\bm{x}) = \dfrac{\sigma(\bm{x})}{\sum\limits_{i=1}^{n} \sigma(x_i) + \epsilon}
\label{eq:normsiggattf}
\end{equation}

In Equation~\eqref{eq:normsiggattf}, $\bm{x} \in \mathbb{R}^n$ is a vector and $\epsilon$ is a very small positive real number used to avoid overflow. 

\subsubsection{Optional deletion of positional encodings}

Lastly, we turn the addition of positional encodings at the inputs into an optional operation and perform experiments with and without these abstract representations as it is unclear how important ordering information is for the task and data at hand. 

\input{avtrans}

\section{Experimental setup}
\label{sect:expset}

In this section, we expand upon the experimental setup. We provide details on the audiovisual data set used to train the multimodal transformers and discuss implementation details such as the used data preprocessing and regularization methods. 

\subsection{Data set}

The large-scale multi-label data set for task 4 of DCASE 2017 consists of 51172 training, 488 validation and 1103 test audio samples originated from YouTube. Most of these audio files are 10 seconds long, some are shorter. They are weakly labeled: only clip-level annotations are provided, indicating the presence or absence of 17 audio event classes, which all represent environmental sounds. Temporal knowledge is not included, meaning that there is no information about the time boundaries of the sounds \cite{DCASE2017challenge}. 

Taking an approach similar to a procedure used in prior work, YouTube downloads allow us to readily append visual data to this audio data set. Unfortunately, due to availability issues, videos could not be obtained for all of the audio files in the original data set. Ultimately, we ended up creating a slightly reduced collection of audio and video files consisting of 48883 training, 465 validation and 1050 test samples: about 5\% of samples of the original data was discarded in the process \cite{weakly}.

\subsection{Implementation details}

The audiovisual transformers are built using the TensorFlow toolkit. In this subsection, some important implementation details are elaborated upon \cite{TensorFlow}.

\subsubsection{Data preprocessing}

To obtain auditory frame embeddings, first logarithmic mel spectrograms are extracted from the audio samples. Next, a sliding window is applied to these spectrograms to divide the signal into frames. The length of this window and its hop length are set to 960 ms. The spectral frames are passed through vggish, a model for video tag classification using audio, pretrained using a preliminary version of the YouTube-8M data set. The 128-dimensional outputs of the last feedforward layer of this deep convolutional network are used as frame embeddings \cite{vggish, YouTube8M}.

Visual frame embeddings are obtained in the following way: firstly, still images are sampled from the videos. The sampling times are chosen to coincide with the centers of the possible positions of the sliding window used in the audio preprocessing described above. 
The resulting video frames are then fed into VGG16, a deep convolutional model for image classification, pretrained on the ImageNet data set. The 4096-dimensional outputs of the last feedforward layer of this neural network are used as visual frame embeddings \cite{ImageNet, VGG16}.

Most of the preprocessing steps described above are included in the TensorFlow toolkit and are therefore easy to implement \cite{TensorFlow}.

\subsubsection{Data sampling}

The audiovisual data used for training is strongly unbalanced: the ratio of number of samples containing a sound of the most common event class to the number of examples comprising the least likely event is approximately equal to 130. 

To alleviate this issue, we use a sampling scheme proposed in prior work: for each training batch, class labels are non-uniformly sampled and appropriate data examples are fetched. This happens in a way that limits the maximum class imbalance ratio to 5 \cite{Surrey}.

We train the models for a maximum of 30 epochs. In each training epoch, 300 batches consisting of 40 data samples are presented to the networks, resulting in a total of 12000 examples per epoch.

\subsubsection{Number of encoder and decoder blocks}

The number of encoder and decoder blocks is referred to with the letter N in Figure~\ref{fig:avtrans}. We set this hyperparameter equal to three.

\subsubsection{Number of attention heads}

We use three attention heads in all multi-head attention blocks in both the encoder and decoder of the audiovisual transformers.

\subsubsection{Feedforward ReLU subnetworks}

The feedforward ReLU subnetworks referenced in Figure~\ref{fig:avtrans} match the ones used in the original transformer shown in Figure~\ref{fig:mttrans}: each subnet consists of one ReLU layer followed by a linear layer without activation function \cite{transformers}.

\subsubsection{Number of neurons}

All feedforward layers with and without activation functions, including all linear transformations used, consist of 128 neurons. 

\subsubsection{Regularization methods}

Identically to the original transformer architecture, computation of the output of each multi-head attention block of the audiovisual transformer is followed by addition with its corresponding residual connection and layer normalization, a popular regularization technique which performs sample-wise scaling. We also apply dropout with a rate of 0.1 to the outputs of the following parts of the model \cite{transformers, residual, layer, dropout}:

\begin{itemize}
    \item Linear layers at the input of the encoder and decoder
    \item Attention function in the multi-head attention blocks
    \item Final linear map in the multi-head attention blocks
    \item Both layers of the feedforward ReLU subnetworks
\end{itemize}

\subsubsection{Loss function and optimization algorithm}

We use the well-known binary cross entropy loss function in conjunction with the Adam optimization algorithm (using standard settings as defined in the TensorFlow toolkit) to train the transformers \cite{adam, TensorFlow}.

Early stopping is applied if at any point during training the cross entropy calculated on the validation data is higher than all validation losses obtained after the last 7 epochs. 

\subsubsection{Epoch-level ensembling}

Instead of only using the model obtained after the final training epoch to obtain audio event class probabilities, we employ the epoch-level ensembling method proposed in the original work on transformers: after each of the last 7 training epochs the model is saved. Eventually, these saved states are used to obtain 7 sets of probabilities, which are then averaged class-wise to obtain a more robust output \cite{transformers}.

\subsubsection{Prediction thresholding}

The final goal of the audiovisual transformer models is to make binary decisions: they should predict whether or not certain audio event classes are present in a given clip. The continuous probabilities at the output of the transformer systems thus should be converted into binary values via thresholding. Preliminary experiments have shown that using a global threshold which maximizes the performance of the models on the validation data is favorable to utilizing class-dependent optimal thresholds, hence the former approach is taken.

\section{Experimental results}
\label{sect:results}

In this section, we discuss the results we obtained by training the audiovisual transformer neural networks. We split this discussion into two parts: the first is the quantitative analysis, where the performance of the models is numerically evaluated and compared to other approaches, the second part is the qualitative analysis which provides interesting, interpretable insights into the workings of the audiovisual transformers. 

\input{f1_avtrans}

\subsection{Quantitative analysis}

We quantitatively evaluate the audiovisual transformer models by measuring micro-averaged F1 scores on the test samples of the employed data. This metric is chosen as it is very often utilized in the field of audio event recognition, e.g., in task 4 of the DCASE 2017 challenge, where the used data originates from \cite{DCASE2017challenge, metrics}.

We report F1 scores which are averaged over 25 training runs to ensure the results are reliable. This is practically feasible as a consequence of the use of fixed pretrained embeddings and due to the fact that transformers are very quick to train, in contrast to neural models containing recurrent connections.

Table~\ref{tab:f1} contains the measured F1 scores on the audiovisual data set created in this work which is described in Section~\ref{sect:expset}. The referred settings are described in detail in Section~\ref{sect:multtrans}. This table contains results for multimodal transformers employing both audio and video as well as transformers utilizing only a single modality. For the models using only audio, performance on the slightly larger (complete) data set of task 4 of the DCASE 2017 challenge also mentioned in Section~\ref{sect:expset} is reported between parentheses \cite{DCASE2017challenge}.

In the rest of this section, we analyze these numerical results and compare to approaches presented in prior works.

\subsubsection{Evaluation of positional encodings}

The unimodal video transformers benefit from the utilization of positional encodings, albeit only slightly. However, employing these abstract position representations is strongly detrimental to the models utilizing only audio. 

The micro-averaged F1 scores obtained by multimodal transformers employing both audio and video show a similar trend: adding positional encodings to the used visual frame embeddings provides minor improvements, adding these abstract position representations to both the auditory and visual embeddings causes a severe deterioration in performance.

\subsubsection{Evaluation of aggegration methods and attention functions}
\label{subsubsect:att}

Generally, audiovisual transformers employing mean pooling as aggregation method seem to perform slightly better than those using max pooling. The disparity in F1 scores is the most pronounced for the unimodal models utilizing only audio. 

Remarkably, transformers utilizing different types of attention functions perform very similarly. In particular, we note that there is no meaningful difference in F1 scores between models employing normalized (softmax, normalized sigmoid) and unnormalized attention functions (sigmoid). 

\subsubsection{Comparison to prior work in audio classification}

We list relevant prior audio event classification results measured on the test samples of the data set of task 4 of the DCASE 2017 challenge in Table~\ref{tab:f1_aural}. These micro-averaged F1 scores can most directly be compared to the outcomes reported between parentheses in Table~\ref{tab:f1}, as their computation is based on the same data \cite{DCASE2017challenge}.

\input{f1_aural}

The transformers outdo the models in Table~\ref{tab:f1_aural}. We speculate that this can be attributed to the power of the multi-head attention mechanism and the fact that they employ rich, externally pretrained embeddings instead of training audio features from scratch. 

\subsubsection{Comparison to prior work in audiovisual classification}

\input{v2a_1}

The multimodal transformers employing both audio and video can be fairly compared to the two-stream audiovisual neural network introduced in prior work: this model performs environmental audio event classification based on audiovisual data acquired in the same manner as the procedure used in this work. The micro-averaged F1 score achieved by this model is reported in Table~\ref{tab:f1_audiovisual}.

\input{f1_audiovisual}

The multimodal audiovisual transformers strongly improve upon the two-stream audiovisual neural network. As both of these types of models use nearly the same data and employ similar externally pretrained embeddings, this truly illustrates the power of the transformer architecture for the task at hand. 

As can be inferred from Table~\ref{tab:f1}, the best performing multimodal transformer type achieves a micro-averaged F1 score of 70.1\%. To the best of our knowledge, this is the best classification result achieved for the considered data.

\subsubsection{Comparison of unimodal and multimodal transformers}

A comparison between the transformers using both audio and video and the transformers utilizing only audio unveils the usefulness of visual information for environmental audio event classification: on average, there is a difference in F1 scores of about 10\%.

Initially, it is reasonable to expect unimodal transformers employing visual information to perform worse than those utilizing audio, since the annotation of the data at hand only indicates the presence or absence of audio event classes as outlined in Section~\ref{sect:expset}. Additionally, the used audiovisual data set was created by simply extending labeled audio files originated from YouTube with their corresponding videos, as also discussed in Section~\ref{sect:expset}, and therefore there is no guarantee that the appended visual material contains valuable information. By conducting a manual inspection of the data set, we were even able to confirm that not all of the downloaded videos are informative for the considered classification task. Despite these remarks, we still find that the unimodal visual models slightly outperform the unimodal models based on audio. 

This peculiar observation can be explained by the fact that in many cases, images or videos are more useful than auditory signals. For example, multiple classes in the used data set pertain to similar vehicle sounds such as police car, fire truck and ambulance sirens: even for humans, differentiating these events solely based on audio can be quite difficult and visual information may be of great help. Another plausible reason for this strange outcome is a disparity in richness of the used pretrained auditory and visual embeddings, which are described in Section~\ref{sect:expset}.

\subsection{Qualitative analysis}
\label{subsect:qual}

\input{v2a_2}

\input{v2a_3}

\input{a2v}

We qualitatively analyze the transformers utilizing both audio and video by visualizing cross-modal attention weights in a manner similar to a procedure followed in prior work, to investigate the ability of these models to perform multimodal synchronization \cite{transformers}. 

\subsubsection{Models using video/audio as first/second inputs respectively}

Figures~\ref{fig:v2a}, \ref{fig:v2a_ex_1} and \ref{fig:v2a_ex_2} contain visualizations for transformers without positional encodings which utilize video as first and audio as second input, mean pooling for aggregation and employ the softmax attention function. A full description of these configurations can be found in Section~\ref{sect:multtrans}. Each figure contains a number of auditory (spectral) and visual frames for a test sample of the audiovisual data set described in Section~\ref{sect:expset}, as well as a heat map of the cross-modal attention weights obtained for the circled audio query frame from the cross-modal multi-head attention component in the last decoder block of the corresponding model. Each row in each heat map pertains to a different attention head.

In all of the cases visualized in Figures~\ref{fig:v2a}, \ref{fig:v2a_ex_1} and \ref{fig:v2a_ex_2}, the models are able to detect visual frames relevant to the environmental sounds present in the circled audio query frames. 

For models using video as first input and audio as second input, we were able to discover similarly interesting attention patterns for many other validation and test samples. Strangely, this was only possible for audiovisual transformers employing the softmax attention function: the architectures utilizing the sigmoid and normalized sigmoid attention functions did not allow for meaningful visualizations, even though they achieve comparable performance as reported in Section~\ref{subsubsect:att}. 

\subsubsection{Models using audio/video as first/second inputs respectively}

In Figure~\ref{fig:a2v}, we present an example for an audiovisual transformer with visual positional encodings which utilizes audio and video as first and second input modalities respectively, mean pooling as aggregation method and uses softmax attention. The figure contains 10 (spectral) auditory and visual frames for a validation sample of the data set described in Section~\ref{sect:expset}, as well as a heat map of the attention weights obtained for the circled video query frame from the cross-modal multi-head attention component in the last decoder block of the associated model. Each row in the heat map pertains to a different attention head.

The circled video query frame of the sample visualized in Figure~\ref{fig:a2v} is a picture of someone skateboarding. Through its attention heads, the corresponding model is able to detect the relevant auditory frames containing skateboarding sounds, which are circled.

For this type of transformer, discovering synchronization patterns in the attention weights proved to be quite difficult. This can likely be explained by the fact that sounds are naturally much less localized compared to visual events. Indeed, also for humans, pointing to specific audio frames as containing valuable information is much harder than pointing to specific video frames, i.e., images.

\section{Conclusion}
\label{sect:concl}

We tackled the task of environmental event classification by drawing inspiration from machine translation: we changed the popular transformer architecture into a model which can use auditory as well as visual data to compute multi-label audio event predictions.

We experimented with these transformers on an audiovisual data set, obtained by extending an existing weakly labeled auditory data set with visual information. The used multi-label data contains clip-level annotation indicating the presence of 17 classes of environmental sounds, without disclosure of temporal information. 

We showed that the proposed models strongly improve upon prior approaches in the fields of audio as well as audiovisual classification and achieve state-of-the-art results.

We also proved the usefulness of visual information for the task at hand, i.e., audio event recognition. This shows that it certainly can be worth researching multimodal audiovisual models.

In addition, we visualized internal attention patterns of the audiovisual transformers. This way, we were able to demonstrate that they show potential for multimodal synchronization.

In the future, we will use these transformers to tackle some other audiovisual data collections and learn more about their properties. In particular, we are keen to analyze the impact of scale: we will experiment with the small-scale data set of task 4 of DCASE 2018, and Audio Set, which is humongous in size. In addition, we will look into how the models can be used for unsupervised tasks \cite{DCASE2018proceedings, AudioSet}.

\begin{acks}
This work is supported by a PhD Fellowship of Research Foundation Flanders (FWO-Vlaanderen).
\end{acks}

\setcitestyle{nosort}
\bibliographystyle{ACM-Reference-Format}
\bibliography{bibliography}

\end{document}

%% file: mttrans.tex
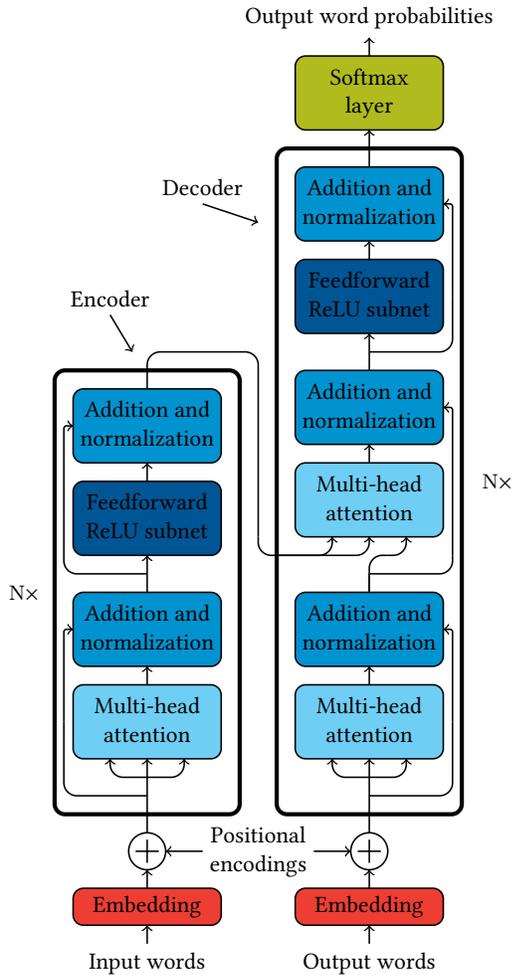
\begin{figure}[!ht]
\centering
\resizebox{0.405\textwidth}{!}{
\begin{tikzpicture}[semithick]
\node [black, below, font=\normalsize] at (1, 0) {Input words};
\node [black, below, font=\normalsize] at (4, 0) {Output words};
\node [black, font=\normalsize] at (2.5, 1.25) {
\begin{tabular}{c}
Positional \\
encodings
\end{tabular}
};
\draw [black, ->] (1.75, 1.25) -- (1.25, 1.25);
\draw [black, ->] (3.25, 1.25) -- (3.75, 1.25);
\draw [black, ->] (1,0) -- (1,0.25);
\draw [black, ->] (4,0) -- (4,0.25);
\draw (0,0.25) [black, rounded corners, fill=\colortwo] rectangle (2, 0.75) node[midway] {Embedding};
\draw (3,0.25) [black, rounded corners, fill=\colortwo] rectangle (5, 0.75) node[midway] {Embedding};
\draw [black, opacity=0] (5,1) -- (6,1);
\draw [black, ->] (1,0.75) -- (1,1);
\draw [black, ->] (4,0.75) -- (4,1);
\draw (1,1.25) node[cross=3.75, rotate=45] {};
\draw (1,1.25) circle (0.25);
\draw (4,1.25) node[cross=3.75, rotate=45] {};
\draw (4,1.25) circle (0.25);
\draw (-0.25,1.75) [ultra thick, black, rounded corners] rectangle (2.25, 7.75);
\node [black, left] at (-0.35, 4.75) {N$\times$};
\draw (2.75,1.75) [ultra thick, black, rounded corners] rectangle (5.25, 10.75);
\node [black, right] at (5.25, 6.25) {N$\times$};
\draw [black, ->] (1,1.5) -- (1,2.5);
\draw [black, ->, rounded corners=5] (1,2.25) -| (0.5,2.5);
\draw [black, ->, rounded corners=5] (1,2.25) -| (1.5,2.5);
\draw [black, ->] (4,1.5) -- (4,2.5);
\draw [black, ->, rounded corners=5] (4,2.25) -| (3.5,2.5);
\draw [black, ->, rounded corners=5] (4,2.25) -| (4.5,2.5);
\draw (0,2.5) [black, rounded corners, fill=\colorfour] rectangle (2, 3.5) node[midway] {
\begin{tabular}{c}
Multi-head \\
attention
\end{tabular}
};
\draw (3,2.5) [black, rounded corners, fill=\colorfour] rectangle (5, 3.5) node[midway] {
\begin{tabular}{c}
Multi-head \\
attention
\end{tabular}
};
\draw [black, ->] (1,3.5) -- (1,3.75);
\draw (0,3.75) [black, rounded corners, fill=\colorfive] rectangle (2, 4.75) node[midway] {
\begin{tabular}{c}
Addition and \\
normalization
\end{tabular}
};
\draw [black, rounded corners=5] (1,2) -| (-0.125,3);
\draw [black, ->, rounded corners=1] (-0.125,3) |- (0,4.25);
\draw [black, ->] (4,3.5) -- (4,3.75);
\draw (3,3.75) [black, rounded corners, fill=\colorfive] rectangle (5, 4.75) node[midway] {
\begin{tabular}{c}
Addition and \\
normalization
\end{tabular}
};
\draw [black, rounded corners=5] (4,2) -| (5.125,3);
\draw [black, ->, rounded corners=1] (5.125,3) |- (5,4.25);
\draw [black, ->] (1,4.75) -- (1,5.25);
\draw [black] (4,4.75) -- (4,5);
\draw [black, rounded corners=5] (4,4.75) |- (4.25,5.25);
\draw [black, ->, rounded corners=5] (4.25,5.25) -| (4.5,5.5);
\draw (0,5.25) [black, rounded corners, fill=\colorthree] rectangle (2, 6.25) node[midway] {
\begin{tabular}{c}
Feedforward \\
ReLU subnet
\end{tabular}
};
\draw (3,5.5) [black, rounded corners, fill=\colorfour] rectangle (5, 6.5) node[midway] {
\begin{tabular}{c}
Multi-head \\
attention
\end{tabular}
};
\draw [black, ->] (1,6.25) -- (1,6.5);
\draw (0,6.5) [black, rounded corners, fill=\colorfive] rectangle (2, 7.5) node[midway] {
\begin{tabular}{c}
Addition and \\
normalization
\end{tabular}
};
\draw [black, rounded corners=5] (1,5) -| (-0.125,5.75);
\draw [black, ->, rounded corners=1] (-0.125,5.75) |- (0,7);
\draw [black, ->] (4,6.5) -- (4,6.75);
\draw (3,6.75) [black, rounded corners, fill=\colorfive] rectangle (5, 7.75) node[midway] {
\begin{tabular}{c}
Addition and \\
normalization
\end{tabular}
};
\draw [black, rounded corners=5] (4,5) -| (5.125,5.25);
\draw [black, ->, rounded corners=1] (5.125,5.25) |- (5,7.25);
\draw [black, rounded corners=5] (1,7.5) |- (2.25,8);
\draw [black, rounded corners=5] (2.25,8) -| (2.5,5.5);
\draw [black, rounded corners=5] (2.5,5.5) |- (2.75,5.25);
\draw [black, ->, rounded corners=5] (2.75,5.25) -| (4,5.5);
\draw [black, ->, rounded corners=5] (2.75,5.25) -| (3.5,5.5);
\draw [black, ->] (4,7.75) -- (4,8.25);
\draw (3,8.25) [black, rounded corners, fill=\colorthree] rectangle (5, 9.25) node[midway] {
\begin{tabular}{c}
Feedforward \\
ReLU subnet
\end{tabular}
};
\draw [black, ->] (4,9.25) -- (4,9.5);
\draw (3,9.5) [black, rounded corners, fill=\colorfive] rectangle (5, 10.5) node[midway] {
\begin{tabular}{c}
Addition and \\
normalization
\end{tabular}
};
\draw [black, rounded corners=5] (4,8) -| (5.125,8.75);
\draw [black, ->, rounded corners=1] (5.125,8.75) |- (5,10);
\draw [black, ->] (4,10.5) -- (4,11);
\draw (3,11) [black, rounded corners, fill=\colornull] rectangle (5, 12) node[midway] {
\begin{tabular}{c}
Softmax \\
layer
\end{tabular}
};
\draw [black, ->] (4,12) -- (4,12.25);
\node [black, above, font=\normalsize] at (4, 12.25) {Output word probabilities};
\node [black, above, font=\normalsize] at (0.5, 8.5) {Encoder};
\draw [black, ->] (0.5, 8.5) -- (0.8,8);
\node [black, above, font=\normalsize] at (1.75, 10) {Decoder};
\draw [black, ->] (1.75, 10) -- (2.5,9.75);
\end{tikzpicture}
}
\caption{Transformer model for machine translation \cite{transformers}}
\Description{A schematic representation of the original transformer model used in the field of machine translation.}
\label{fig:mttrans}
\end{figure}

%% file: mhatt.tex
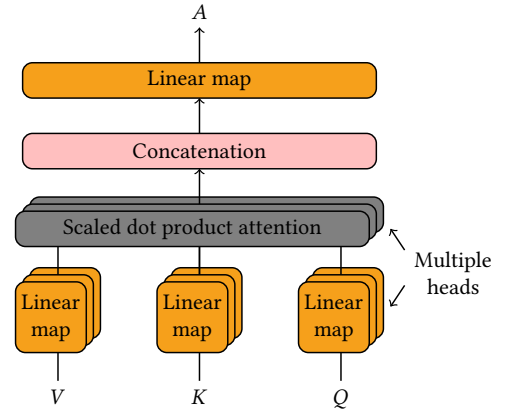
\begin{figure}[!ht]
\centering
\resizebox{0.475\textwidth}{!}{
\begin{tikzpicture}[semithick]
\node [black, below, font=\normalsize] at (1, 0) {$V$};
\node [black, below, font=\normalsize] at (3, 0) {$K$};
\node [black, below, font=\normalsize] at (5, 0) {$Q$};
\draw [black, ->] (1, 0) -- (1, 0.5);
\draw [black, ->] (3, 0) -- (3, 0.5);
\draw [black, ->] (5, 0) -- (5, 0.5);
\draw (0.6, 0.6) [black, rounded corners, fill=\colorseven] rectangle (1.6, 1.6);
\draw (0.5, 0.5) [black, rounded corners, fill=\colorseven] rectangle (1.5, 1.5);
\draw (0.4, 0.4) [black, rounded corners, fill=\colorseven] rectangle (1.4, 1.4) node[midway]{
\begin{tabular}{c}
Linear \\
map
\end{tabular}
};
\draw (2.6, 0.6) [black, rounded corners, fill=\colorseven] rectangle (3.6, 1.6);
\draw (2.5, 0.5) [black, rounded corners, fill=\colorseven] rectangle (3.5, 1.5);
\draw (2.4, 0.4) [black, rounded corners, fill=\colorseven] rectangle (3.4, 1.4) node[midway]{
\begin{tabular}{c}
Linear \\
map
\end{tabular}
};
\draw (4.6, 0.6) [black, rounded corners, fill=\colorseven] rectangle (5.6, 1.6);
\draw (4.5, 0.5) [black, rounded corners, fill=\colorseven] rectangle (5.5, 1.5);
\draw (4.4, 0.4) [black, rounded corners, fill=\colorseven] rectangle (5.4, 1.4) node[midway]{
\begin{tabular}{c}
Linear \\
map
\end{tabular}
};
\draw [black, ->] (1, 1.5) -- (1, 2);
\draw [black, ->] (3, 1.5) -- (3, 2);
\draw [black, ->] (5, 1.5) -- (5, 2);
\draw [black, ->] (5.9, 1.35) -- (5.7, 1.05);
\draw [black, ->] (5.9, 1.85) -- (5.7, 2.15);
\node [black, right, font=\normalsize] at (5.75, 1.5) {
\begin{tabular}{c}
Multiple \\
heads
\end{tabular}
};
\node [black, left, opacity=0, font=\normalsize] at (0.25, 1.5) {
\begin{tabular}{c}
Multiple \\
heads
\end{tabular}
};
\draw [black, ->] (3, 1.5) -- (3, 2);
\draw (0.6, 2.1) [black, rounded corners, fill=\coloreight] rectangle (5.6, 2.6);
\draw (0.5, 2) [black, rounded corners, fill=\coloreight] rectangle (5.5, 2.5);
\draw (0.4, 1.9) [black, rounded corners, fill=\coloreight] rectangle (5.4, 2.4) node[midway]{Scaled dot product attention};
\draw [black, ->] (3, 2.5) -- (3, 3);
\draw (0.5, 3) [black, rounded corners, fill=\colornine] rectangle (5.5, 3.5) node[midway]{Concatenation};
\draw [black, ->] (3, 3.5) -- (3, 4);
\draw (0.5, 4) [black, rounded corners, fill=\colorseven] rectangle (5.5, 4.5) node[midway]{Linear map};
\draw [black, ->] (3, 4.5) -- (3, 5);
\node [black, above, font=\normalsize] at (3, 5) {$A$};
\end{tikzpicture}
}
\caption{Multi-head scaled dot product attention block \cite{transformers}}
\Description{A schematic representation of the multi-head scaled dot product attention block used in the transformer architecture.}
\label{fig:mhatt}
\end{figure}

%% file: avtrans.tex
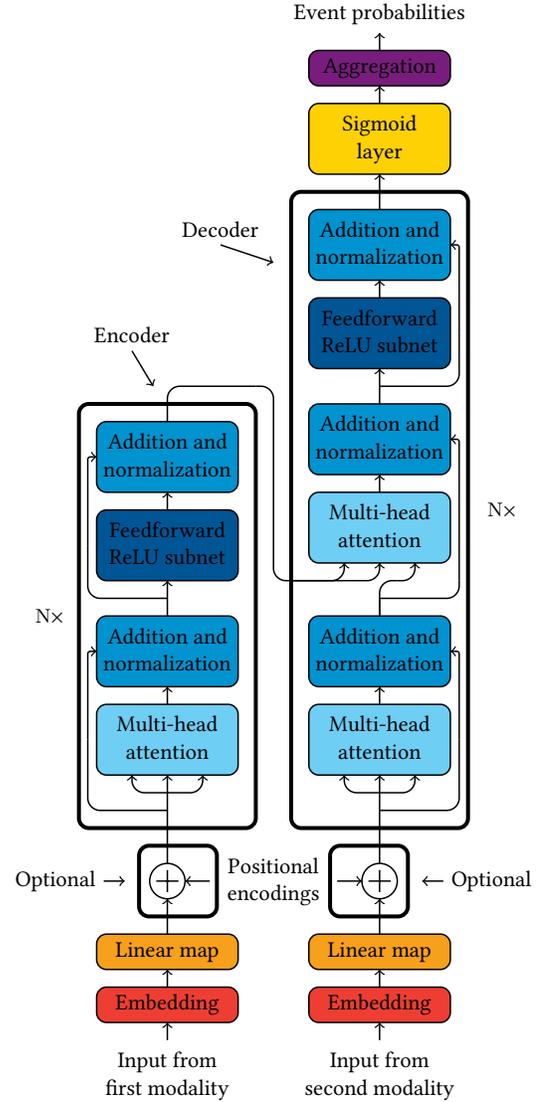
\begin{figure}[!ht]
\centering
\resizebox{0.405\textwidth}{!}{
\begin{tikzpicture}[semithick]
\node [black, below, font=\normalsize] at (1, -1.25) {
\begin{tabular}{c}
Input from  \\
first modality
\end{tabular}
};
\node [black, below, font=\normalsize] at (4, -1.25) {
\begin{tabular}{c}
Input from  \\
second modality
\end{tabular}
};
\draw (0.6,0.5) [ultra thick, black, rounded corners] rectangle (1.7, 1.5);
\draw (3.3,0.5) [ultra thick, black, rounded corners] rectangle (4.4, 1.5);
\node [black, font=\normalsize] at (2.5, 1) {
\begin{tabular}{c}
Positional \\
encodings
\end{tabular}
};
\draw [black, ->] (1.6, 1) -- (1.25, 1);
\draw [black, ->] (3.4, 1) -- (3.75, 1);
\draw [black, ->] (1,-1.25) -- (1,-1);
\draw [black, ->] (4,-1.25) -- (4,-1);
\draw (0,-1) [black, rounded corners, fill=\colortwo] rectangle (2, -0.5) node[midway] {Embedding};
\draw (3,-1) [black, rounded corners, fill=\colortwo] rectangle (5, -0.5) node[midway] {Embedding};
\draw [black, ->] (1,-0.5) -- (1,-0.25);
\draw [black, ->] (4,-0.5) -- (4,-0.25);
\draw (0,-0.25) [black, rounded corners, fill=\colorseven] rectangle (2, 0.25) node[midway] {Linear map};
\draw (3,-0.25) [black, rounded corners, fill=\colorseven] rectangle (5, 0.25) node[midway] {Linear map};
\draw [black, opacity=0] (5,1) -- (6,1);
\draw [black, ->] (1,0.25) -- (1,0.75);
\draw [black, ->] (4,0.25) -- (4,0.75);
\draw (1,1) node[cross=3.75, rotate=45] {};
\draw (1,1) circle (0.25);
\draw (4,1) node[cross=3.75, rotate=45] {};
\draw (4,1) circle (0.25);
\draw (-0.25,1.75) [ultra thick, black, rounded corners] rectangle (2.25, 7.75);
\node [black, left] at (-0.35, 4.75) {N$\times$};
\draw (2.75,1.75) [ultra thick, black, rounded corners] rectangle (5.25, 10.75);
\node [black, right] at (5.25, 6.25) {N$\times$};
\draw [black, ->] (1,1.25) -- (1,2.5);
\draw [black, ->, rounded corners=5] (1,2.25) -| (0.5,2.5);
\draw [black, ->, rounded corners=5] (1,2.25) -| (1.5,2.5);
\draw [black, ->] (4,1.25) -- (4,2.5);
\draw [black, ->, rounded corners=5] (4,2.25) -| (3.5,2.5);
\draw [black, ->, rounded corners=5] (4,2.25) -| (4.5,2.5);
\draw (0,2.5) [black, rounded corners, fill=\colorfour] rectangle (2, 3.5) node[midway] {
\begin{tabular}{c}
Multi-head \\
attention
\end{tabular}
};
\draw (3,2.5) [black, rounded corners, fill=\colorfour] rectangle (5, 3.5) node[midway] {
\begin{tabular}{c}
Multi-head \\
attention
\end{tabular}
};
\draw [black, ->] (1,3.5) -- (1,3.75);
\draw (0,3.75) [black, rounded corners, fill=\colorfive] rectangle (2, 4.75) node[midway] {
\begin{tabular}{c}
Addition and \\
normalization
\end{tabular}
};
\draw [black, rounded corners=5] (1,2) -| (-0.125,3);
\draw [black, ->, rounded corners=1] (-0.125,3) |- (0,4.25);
\draw [black, ->] (4,3.5) -- (4,3.75);
\draw (3,3.75) [black, rounded corners, fill=\colorfive] rectangle (5, 4.75) node[midway] {
\begin{tabular}{c}
Addition and \\
normalization
\end{tabular}
};
\draw [black, rounded corners=5] (4,2) -| (5.125,3);
\draw [black, ->, rounded corners=1] (5.125,3) |- (5,4.25);
\draw [black, ->] (1,4.75) -- (1,5.25);
\draw [black] (4,4.75) -- (4,5);
\draw [black, rounded corners=5] (4,4.75) |- (4.25,5.25);
\draw [black, ->, rounded corners=5] (4.25,5.25) -| (4.5,5.5);
\draw (0,5.25) [black, rounded corners, fill=\colorthree] rectangle (2, 6.25) node[midway] {
\begin{tabular}{c}
Feedforward \\
ReLU subnet
\end{tabular}
};
\draw (3,5.5) [black, rounded corners, fill=\colorfour] rectangle (5, 6.5) node[midway] {
\begin{tabular}{c}
Multi-head \\
attention
\end{tabular}
};
\draw [black, ->] (1,6.25) -- (1,6.5);
\draw (0,6.5) [black, rounded corners, fill=\colorfive] rectangle (2, 7.5) node[midway] {
\begin{tabular}{c}
Addition and \\
normalization
\end{tabular}
};
\draw [black, rounded corners=5] (1,5) -| (-0.125,5.75);
\draw [black, ->, rounded corners=1] (-0.125,5.75) |- (0,7);
\draw [black, ->] (4,6.5) -- (4,6.75);
\draw (3,6.75) [black, rounded corners, fill=\colorfive] rectangle (5, 7.75) node[midway] {
\begin{tabular}{c}
Addition and \\
normalization
\end{tabular}
};
\draw [black, rounded corners=5] (4,5) -| (5.125,5.25);
\draw [black, ->, rounded corners=1] (5.125,5.25) |- (5,7.25);
\draw [black, rounded corners=5] (1,7.5) |- (2.25,8);
\draw [black, rounded corners=5] (2.25,8) -| (2.5,5.5);
\draw [black, rounded corners=5] (2.5,5.5) |- (2.75,5.25);
\draw [black, ->, rounded corners=5] (2.75,5.25) -| (4,5.5);
\draw [black, ->, rounded corners=5] (2.75,5.25) -| (3.5,5.5);
\draw [black, ->] (4,7.75) -- (4,8.25);
\draw (3,8.25) [black, rounded corners, fill=\colorthree] rectangle (5, 9.25) node[midway] {
\begin{tabular}{c}
Feedforward \\
ReLU subnet
\end{tabular}
};
\draw [black, ->] (4,9.25) -- (4,9.5);
\draw (3,9.5) [black, rounded corners, fill=\colorfive] rectangle (5, 10.5) node[midway] {
\begin{tabular}{c}
Addition and \\
normalization
\end{tabular}
};
\draw [black, rounded corners=5] (4,8) -| (5.125,8.75);
\draw [black, ->, rounded corners=1] (5.125,8.75) |- (5,10);
\draw [black, ->] (4,10.5) -- (4,11);
\draw (3,11) [black, rounded corners, fill=\colorone] rectangle (5, 12) node[midway] {
\begin{tabular}{c}
Sigmoid \\
layer
\end{tabular}
};
\draw [black, ->] (4,12) -- (4,12.25);
\draw (3,12.25) [black, rounded corners, fill=\colorsix] rectangle (5, 12.75) node[midway] {Aggregation};
\draw [black, ->] (4,12.75) -- (4,13);
\node [black, above, font=\normalsize] at (4, 13) {Event probabilities};
\node [black, above, font=\normalsize] at (0.5, 8.5) {Encoder};
\draw [black, ->] (0.5, 8.5) -- (0.8,8);
\node [black, above, font=\normalsize] at (1.75, 10) {Decoder};
\draw [black, ->] (1.75, 10) -- (2.5,9.75);
\node [black, right, font=\normalsize] at (4.9, 1) {Optional};
\draw [black, ->] (4.9, 1) -- (4.6, 1);
\node [black, left, font=\normalsize] at (0.1, 1) {Optional};
\draw [black, ->] (0.1, 1) -- (0.4, 1);
\end{tikzpicture}
}
\caption{Audiovisual transformer model}
\Description{A schematic representation of the adapted audiovisual transformer model}
\label{fig:avtrans}
\end{figure}

%% file: f1_avtrans.tex
\begin{table*}[!ht]
\caption{F1 scores of audiovisual transformer models}
\label{tab:f1}
\begin{tabular}{@{}lllcccccc@{}}
\toprule
\multirow{5}[3]{*}{\begin{tabular}{@{}l@{}} modality of \\ first input \end{tabular}} & \multirow{5}[3]{*}{\begin{tabular}{@{}l@{}} modality of \\ second input \end{tabular}} & \multirow{5}[3]{*}{\begin{tabular}{@{}l@{}} modalities \\ utilizing \\ positional \\ encodings \end{tabular}} & \multicolumn{6}{c}{F1 score} \\ 
\cmidrule{4-9}
 & & & \multicolumn{3}{c}{mean pooling as aggregation method} & \multicolumn{3}{c}{max pooling as aggregation method} \\
\cmidrule{4-9}
& & & \multirow{3}{*}{\begin{tabular}{@{}c@{}} softmax \\ attention \end{tabular}} & \multirow{3}{*}{\begin{tabular}{@{}c@{}} sigmoid \\ attention \end{tabular}} & \multirow{3}{*}{\begin{tabular}{@{}c@{}} normalized \\ sigmoid \\ attention \end{tabular}} & \multirow{3}{*}{\begin{tabular}{@{}c@{}} softmax \\ attention \end{tabular}} & \multirow{3}{*}{\begin{tabular}{@{}c@{}} sigmoid \\ attention \end{tabular}} & \multirow{3}{*}{\begin{tabular}{@{}c@{}} normalized \\ sigmoid \\ attention \end{tabular}} \\
& & & & & & & & \\
& & & & & & & & \\
\midrule
\noalign{\vskip-2.5pt}
\multicolumn{1}{@{}>{\columncolor{gray!20}[0pt][2.1\tabcolsep]}l}{} & \multicolumn{1}{>{\columncolor{gray!20}[0pt][2.1\tabcolsep]}l}{} & \multicolumn{1}{>{\columncolor{gray!20}[0pt][2.1\tabcolsep]}l}{audio/video} & \multicolumn{1}{>{\columncolor{gray!20}[0pt][2.1\tabcolsep]}c}{66.1\%} & \multicolumn{1}{>{\columncolor{gray!20}[0pt][2.1\tabcolsep]}c}{66.0\%} & \multicolumn{1}{>{\columncolor{gray!20}[0pt][2.1\tabcolsep]}c}{66.3\%} & \multicolumn{1}{>{\columncolor{gray!20}[0pt][2.1\tabcolsep]}c}{66.2\%} & \multicolumn{1}{>{\columncolor{gray!20}[0pt][2.1\tabcolsep]}c}{66.3\%} & \multicolumn{1}{>{\columncolor{gray!20}[0pt][0pt]}c@{}}{\underline{66.4\%}} \\
\noalign{\vskip-0.1pt}
\multicolumn{1}{@{}>{\columncolor{gray!20}[0pt][2.1\tabcolsep]}l}{} & \multicolumn{1}{>{\columncolor{gray!20}[0pt][2.1\tabcolsep]}l}{} & \multicolumn{1}{>{\columncolor{gray!20}[0pt][2.1\tabcolsep]}l}{video} & \multicolumn{1}{>{\columncolor{gray!20}[0pt][2.1\tabcolsep]}c}{69.5\%} & \multicolumn{1}{>{\columncolor{gray!20}[0pt][2.1\tabcolsep]}c}{\underline{69.8\%}} & \multicolumn{1}{>{\columncolor{gray!20}[0pt][2.1\tabcolsep]}c}{69.4\%} & \multicolumn{1}{>{\columncolor{gray!20}[0pt][2.1\tabcolsep]}c}{69.3\%} & \multicolumn{1}{>{\columncolor{gray!20}[0pt][2.1\tabcolsep]}c}{69.5\%} & \multicolumn{1}{>{\columncolor{gray!20}[0pt][0pt]}c@{}}{69.4\%} \\
\noalign{\vskip-0.1pt}
\multicolumn{1}{@{}>{\columncolor{gray!20}[0pt][2.1\tabcolsep]}l}{\multirow{-3}{*}{audio}} & \multicolumn{1}{>{\columncolor{gray!20}[0pt][2.1\tabcolsep]}l}{\multirow{-3}{*}{video}} & \multicolumn{1}{>{\columncolor{gray!20}[0pt][2.1\tabcolsep]}l}{none} & \multicolumn{1}{>{\columncolor{gray!20}[0pt][2.1\tabcolsep]}c}{69.2\%} & \multicolumn{1}{>{\columncolor{gray!20}[0pt][2.1\tabcolsep]}c}{\underline{69.4\%}} & \multicolumn{1}{>{\columncolor{gray!20}[0pt][2.1\tabcolsep]}c}{69.3\%} & \multicolumn{1}{>{\columncolor{gray!20}[0pt][2.1\tabcolsep]}c}{69.1\%} & \multicolumn{1}{>{\columncolor{gray!20}[0pt][2.1\tabcolsep]}c}{69.3\%} & \multicolumn{1}{>{\columncolor{gray!20}[0pt][0pt]}c@{}}{68.9\%} \\
\noalign{\vskip-0.1pt}
\multicolumn{1}{@{}>{\columncolor{gray!5}[0pt][2.1\tabcolsep]}l}{} & \multicolumn{1}{>{\columncolor{gray!5}[0pt][2.1\tabcolsep]}l}{} & \multicolumn{1}{>{\columncolor{gray!5}[0pt][2.1\tabcolsep]}l}{audio/video} & \multicolumn{1}{>{\columncolor{gray!5}[0pt][2.1\tabcolsep]}c}{67.7\%} & \multicolumn{1}{>{\columncolor{gray!5}[0pt][2.1\tabcolsep]}c}{67.5\%} & \multicolumn{1}{>{\columncolor{gray!5}[0pt][2.1\tabcolsep]}c}{\underline{67.8\%}} & \multicolumn{1}{>{\columncolor{gray!5}[0pt][2.1\tabcolsep]}c}{67.3\%} & \multicolumn{1}{>{\columncolor{gray!5}[0pt][2.1\tabcolsep]}c}{67.1\%} & \multicolumn{1}{>{\columncolor{gray!5}[0pt][0pt]}c@{}}{67.6\%} \\
 \noalign{\vskip-0.1pt}
\multicolumn{1}{@{}>{\columncolor{gray!5}[0pt][2.1\tabcolsep]}l}{} & \multicolumn{1}{>{\columncolor{gray!5}[0pt][2.1\tabcolsep]}l}{} & \multicolumn{1}{>{\columncolor{gray!5}[0pt][2.1\tabcolsep]}l}{video} & \multicolumn{1}{>{\columncolor{gray!5}[0pt][2.1\tabcolsep]}c}{69.7\%} & \multicolumn{1}{>{\columncolor{gray!5}[0pt][2.1\tabcolsep]}c}{\underline{70.1\%}} & \multicolumn{1}{>{\columncolor{gray!5}[0pt][2.1\tabcolsep]}c}{69.7\%} & \multicolumn{1}{>{\columncolor{gray!5}[0pt][2.1\tabcolsep]}c}{69.7\%} & \multicolumn{1}{>{\columncolor{gray!5}[0pt][2.1\tabcolsep]}c}{69.8\%} & \multicolumn{1}{>{\columncolor{gray!5}[0pt][0pt]}c@{}}{69.7\%} \\
 \noalign{\vskip-0.1pt}
\multicolumn{1}{@{}>{\columncolor{gray!5}[0pt][2.1\tabcolsep]}l}{\multirow{-3}{*}{video}} & \multicolumn{1}{>{\columncolor{gray!5}[0pt][2.1\tabcolsep]}l}{\multirow{-3}{*}{audio}} & \multicolumn{1}{>{\columncolor{gray!5}[0pt][2.1\tabcolsep]}l}{none} & \multicolumn{1}{>{\columncolor{gray!5}[0pt][2.1\tabcolsep]}c}{69.6\%} & \multicolumn{1}{>{\columncolor{gray!5}[0pt][2.1\tabcolsep]}c}{\underline{69.8\%}} & \multicolumn{1}{>{\columncolor{gray!5}[0pt][2.1\tabcolsep]}c}{69.5\%} & \multicolumn{1}{>{\columncolor{gray!5}[0pt][2.1\tabcolsep]}c}{69.7\%} & \multicolumn{1}{>{\columncolor{gray!5}[0pt][2.1\tabcolsep]}c}{69.7\%} & \multicolumn{1}{>{\columncolor{gray!5}[0pt][0pt]}c@{}}{69.7\%} \\
\noalign{\vskip-0.1pt}
\multicolumn{1}{@{}>{\columncolor{gray!20}[0pt][2.1\tabcolsep]}l}{} & \multicolumn{1}{>{\columncolor{gray!20}[0pt][2.1\tabcolsep]}l}{} & \multicolumn{1}{>{\columncolor{gray!20}[0pt][2.1\tabcolsep]}l}{video} & \multicolumn{1}{>{\columncolor{gray!20}[0pt][2.1\tabcolsep]}c}{59.1\%} & \multicolumn{1}{>{\columncolor{gray!20}[0pt][2.1\tabcolsep]}c}{\underline{59.3\%}} & \multicolumn{1}{>{\columncolor{gray!20}[0pt][2.1\tabcolsep]}c}{\underline{59.3\%}} & \multicolumn{1}{>{\columncolor{gray!20}[0pt][2.1\tabcolsep]}c}{59.0\%} & \multicolumn{1}{>{\columncolor{gray!20}[0pt][2.1\tabcolsep]}c}{59.1\%} & \multicolumn{1}{>{\columncolor{gray!20}[0pt][0pt]}c@{}}{59.0\%} \\
 \noalign{\vskip-0.1pt}
\multicolumn{1}{@{}>{\columncolor{gray!20}[0pt][2.1\tabcolsep]}l}{\multirow{-2}{*}{video}} & \multicolumn{1}{>{\columncolor{gray!20}[0pt][2.1\tabcolsep]}l}{\multirow{-2}{*}{video}} & \multicolumn{1}{>{\columncolor{gray!20}[0pt][2.1\tabcolsep]}l}{none} & \multicolumn{1}{>{\columncolor{gray!20}[0pt][2.1\tabcolsep]}c}{58.8\%} & \multicolumn{1}{>{\columncolor{gray!20}[0pt][2.1\tabcolsep]}c}{\underline{59.1\%}} & \multicolumn{1}{>{\columncolor{gray!20}[0pt][2.1\tabcolsep]}c}{59.0\%} & \multicolumn{1}{>{\columncolor{gray!20}[0pt][2.1\tabcolsep]}c}{58.8\%} & \multicolumn{1}{>{\columncolor{gray!20}[0pt][2.1\tabcolsep]}c}{58.8\%} & \multicolumn{1}{>{\columncolor{gray!20}[0pt][0pt]}c@{}}{58.8\%} \\
\noalign{\vskip-0.1pt}
\multicolumn{1}{@{}>{\columncolor{gray!5}[0pt][2.1\tabcolsep]}l}{} & \multicolumn{1}{>{\columncolor{gray!5}[0pt][2.1\tabcolsep]}l}{} & \multicolumn{1}{>{\columncolor{gray!5}[0pt][2.1\tabcolsep]}l}{audio} & \multicolumn{1}{>{\columncolor{gray!5}[0pt][2.1\tabcolsep]}c}{\underline{57.0\%} (57.0\%)} & \multicolumn{1}{>{\columncolor{gray!5}[0pt][2.1\tabcolsep]}c}{56.9\% (57.0\%)} & \multicolumn{1}{>{\columncolor{gray!5}[0pt][2.1\tabcolsep]}c}{56.8\% (\underline{57.1\%})} & \multicolumn{1}{>{\columncolor{gray!5}[0pt][2.1\tabcolsep]}c}{56.2\% (56.4\%)} & \multicolumn{1}{>{\columncolor{gray!5}[0pt][2.1\tabcolsep]}c}{56.4\% (56.7\%)} & \multicolumn{1}{>{\columncolor{gray!5}[0pt][0pt]}c@{}}{56.4\% (56.7\%)} \\
\noalign{\vskip-0.1pt}
\multicolumn{1}{@{}>{\columncolor{gray!5}[0pt][2.1\tabcolsep]}l}{\multirow{-2}{*}{audio}} & \multicolumn{1}{>{\columncolor{gray!5}[0pt][2.1\tabcolsep]}l}{\multirow{-2}{*}{audio}} & \multicolumn{1}{>{\columncolor{gray!5}[0pt][2.1\tabcolsep]}l}{none} & \multicolumn{1}{>{\columncolor{gray!5}[0pt][2.1\tabcolsep]}c}{58.6\% (59.3\%)} & \multicolumn{1}{>{\columncolor{gray!5}[0pt][2.1\tabcolsep]}c}{\underline{59.1\%} (\underline{59.5\%})} & \multicolumn{1}{>{\columncolor{gray!5}[0pt][2.1\tabcolsep]}c}{58.8\% (59.3\%)} & \multicolumn{1}{>{\columncolor{gray!5}[0pt][2.1\tabcolsep]}c}{58.5\% (59.0\%)} & \multicolumn{1}{>{\columncolor{gray!5}[0pt][2.1\tabcolsep]}c}{58.9\% (59.4\%)} & \multicolumn{1}{>{\columncolor{gray!5}[0pt][0pt]}c@{}}{58.7\% (59.0\%)} \\
\noalign{\vskip-1.7pt}
\bottomrule
\end{tabular}
\end{table*}

%% file: f1_aural.tex
\begin{table}[!ht]
\caption{F1 scores of prior audio classification models}
\label{tab:f1_aural}
\begin{tabular}{@{}lc@{}}
\toprule
model & F1 score\\
\midrule
\noalign{\vskip-2.5pt}
\multicolumn{1}{@{}>{\columncolor{gray!20}[0pt][2.1\tabcolsep]}l}{fusion of gated convolutional recurrent networks \cite{gatedconv}}& \multicolumn{1}{>{\columncolor{gray!20}[0pt][0pt]}c@{}}{55.6\%} \\
\noalign{\vskip-0.1pt}
\multicolumn{1}{@{}>{\columncolor{gray!5}[0pt][2.1\tabcolsep]}l}{capsule-based gated convolutional network \cite{capsule}}& \multicolumn{1}{>{\columncolor{gray!5}[0pt][0pt]}c@{}}{58.6\%} \\
\noalign{\vskip-2.5pt}
\bottomrule
\end{tabular}
\end{table}

%% file: v2a_1.tex
\begin{figure*}[!ht]
\centering
\includegraphics[width=0.785\textwidth]{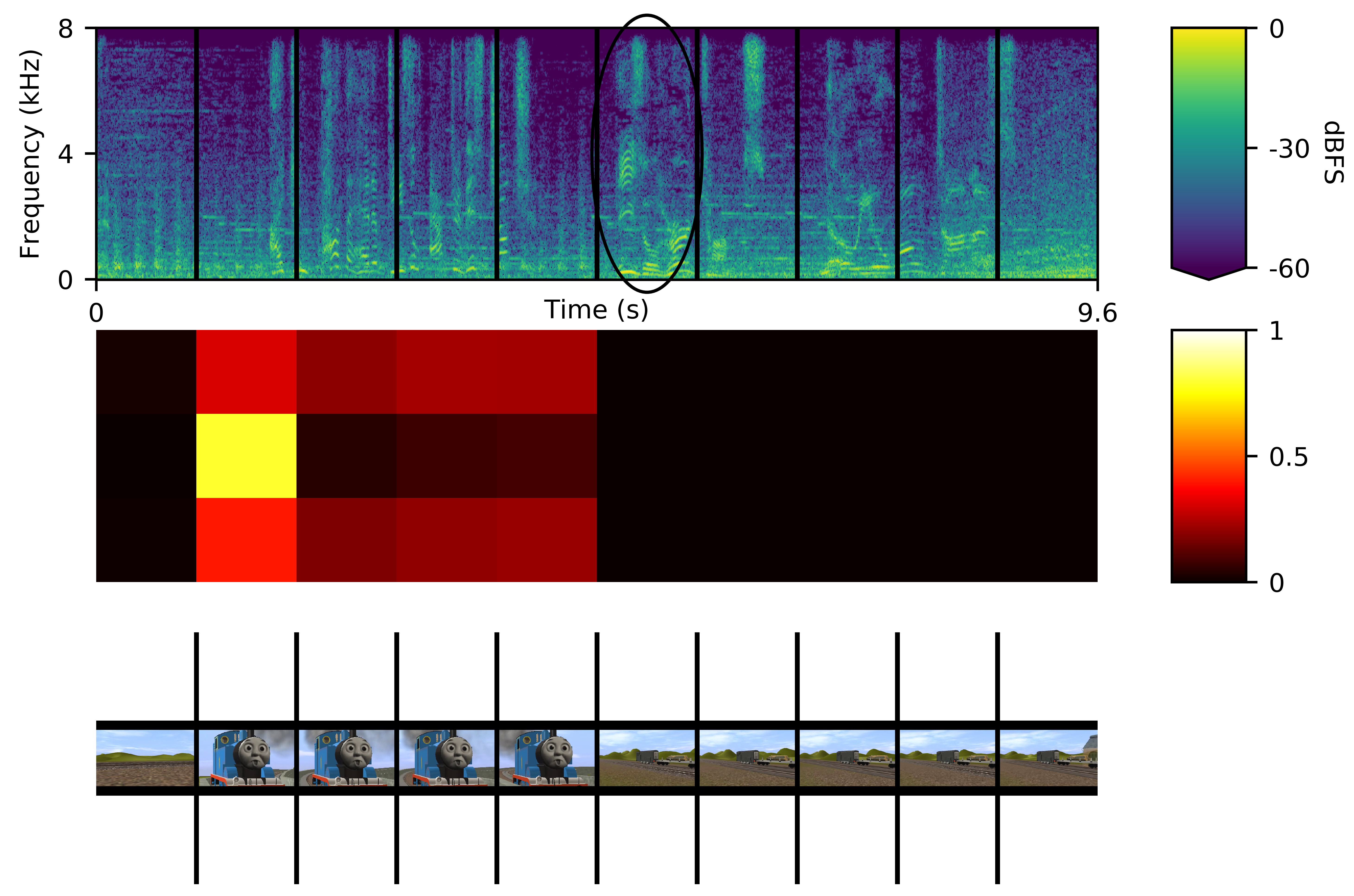}
\caption{Visualization for transformer using video/audio as first/second input modalities (sound produced by train)}
\Description{A visualization of the attention weights for a transformer using video and audio as first and second input modalities respectively. The visualized sample involves the sound produced by a train.}
\label{fig:v2a}
\end{figure*}

%% file: f1_audiovisual.tex
\begin{table}[!ht]
\caption{F1 score of prior audiovisual classification model}
\label{tab:f1_audiovisual}
\begin{tabular}{@{}lc@{}}
\toprule
model & F1 score\\
\midrule
\noalign{\vskip-2.5pt}
\multicolumn{1}{@{}>{\columncolor{gray!20}[0pt][2.1\tabcolsep]}l}{two-stream audiovisual neural network \cite{weakly}}& \multicolumn{1}{>{\columncolor{gray!20}[0pt][0pt]}c@{}}{64.2\%} \\
\noalign{\vskip-2.5pt}
\bottomrule
\end{tabular}
\end{table}

%% file: v2a_2.tex
\begin{figure*}[!ht]
\centering
\includegraphics[width=0.845\textwidth]{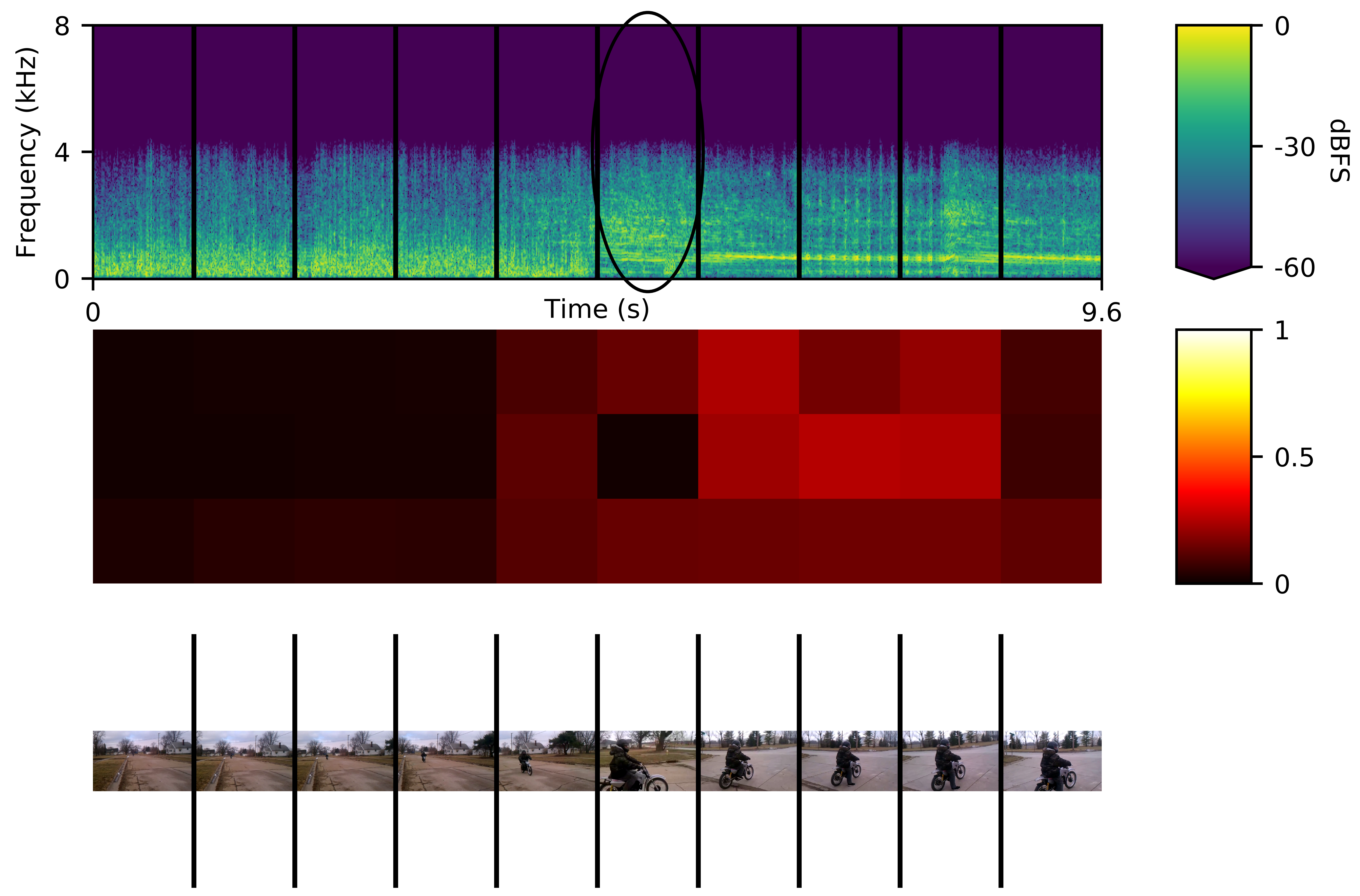}
\caption{Visualization for transformer using video/audio as first/second input modalities (sound produced by motorcycle)}
\Description{A visualization of the attention weights for a transformer using video and audio as first and second input modalities respectively. The visualized sample involves the sound produced by a motorcycle.}
\label{fig:v2a_ex_1}
\end{figure*}

%% file: v2a_3.tex
\begin{figure*}[!ht]
\centering
\includegraphics[width=0.845\textwidth]{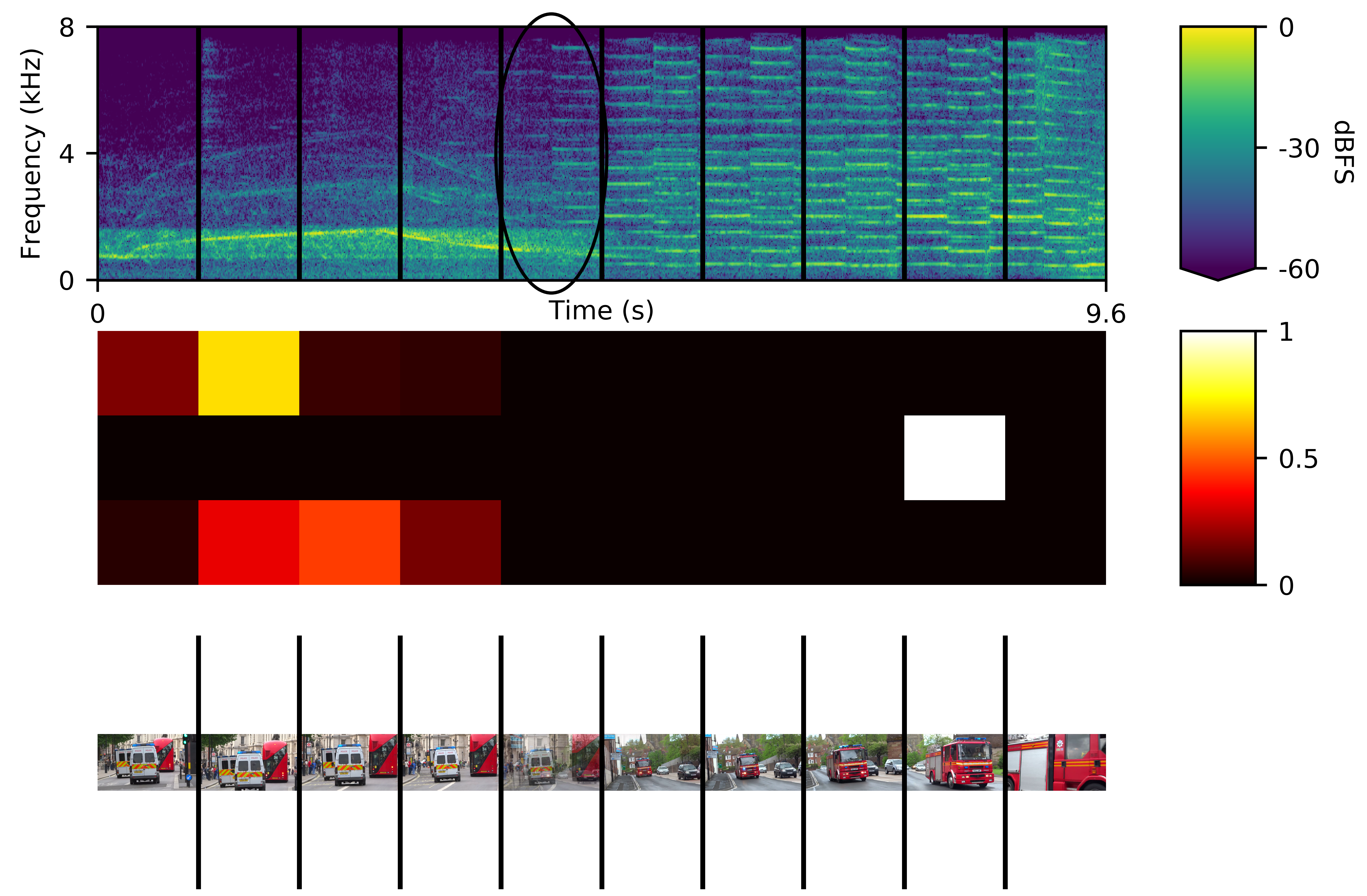}
\caption{Visualization for transformer using video/audio as first/second input modalities (police car and fire truck sirens)}
\Description{A visualization of the attention weights for a transformer using video and audio as first and second input modalities respectively. The visualized sample involves two sounds: a police car and a fire truck siren.}
\label{fig:v2a_ex_2}
\end{figure*}

%% file: a2v.tex
\begin{figure*}[!ht]
\centering
\includegraphics[width=0.755\textwidth]{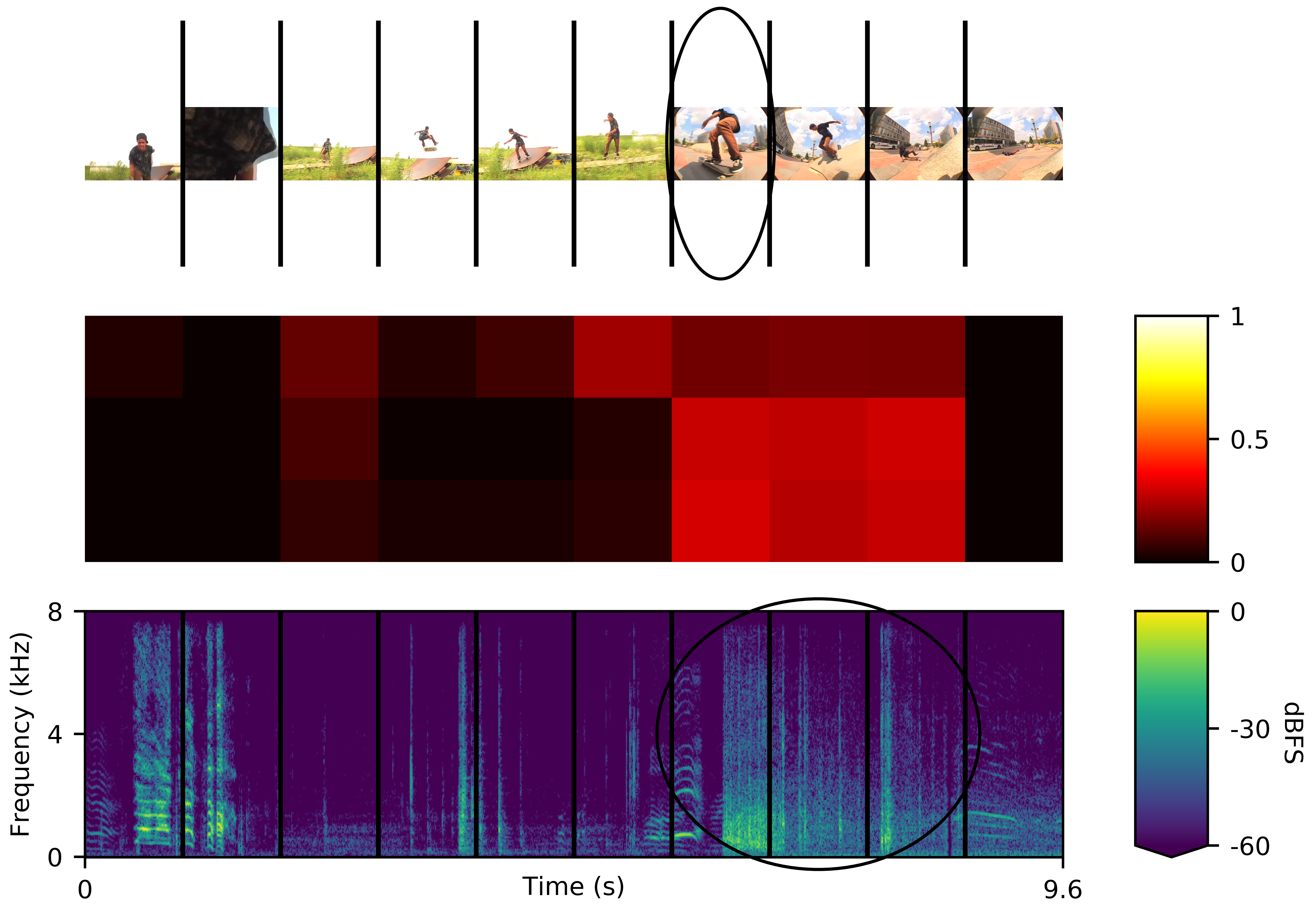}
\caption{Visualization for transformer using audio/video as first/second input modalities (sound produced by skateboard)}
\Description{A visualization of the attention weights for a transformer using audio and video as first and second input modalities respectively. The visualized sample involves the sound produced by a skateboard.}
\label{fig:a2v}
\end{figure*}

%% file: main.bbl

\begin{thebibliography}{21}

%%% the Web address.

\ifx \showCODEN    \undefined \def \showCODEN     #1{\unskip}     \fi
\ifx \showDOI      \undefined \def \showDOI       #1{#1}\fi
\ifx \showISBNx    \undefined \def \showISBNx     #1{\unskip}     \fi
\ifx \showISBNxiii \undefined \def \showISBNxiii  #1{\unskip}     \fi
\ifx \showISSN     \undefined \def \showISSN      #1{\unskip}     \fi
\ifx \showLCCN     \undefined \def \showLCCN      #1{\unskip}     \fi
\ifx \shownote     \undefined \def \shownote      #1{#1}          \fi
\ifx \showarticletitle \undefined \def \showarticletitle #1{#1}   \fi
\ifx \showURL      \undefined \def \showURL       {\relax}        \fi
% invisible to TeX
\providecommand\bibfield[2]{#2}
\providecommand\bibinfo[2]{#2}
\providecommand\natexlab[1]{#1}
\providecommand\showeprint[2][]{arXiv:#2}

\bibitem[\protect\citeauthoryear{Abadi, Barham, Chen, Chen, Davis, Dean, Devin,
  Ghemawat, Irving, Isard, et~al\mbox{.}}{Abadi et~al\mbox{.}}{2016}]%
        {TensorFlow}
\bibfield{author}{\bibinfo{person}{Mart{\'\i}n Abadi}, \bibinfo{person}{Paul
  Barham}, \bibinfo{person}{Jianmin Chen}, \bibinfo{person}{Zhifeng Chen},
  \bibinfo{person}{Andy Davis}, \bibinfo{person}{Jeffrey Dean},
  \bibinfo{person}{Matthieu Devin}, \bibinfo{person}{Sanjay Ghemawat},
  \bibinfo{person}{Geoffrey Irving}, \bibinfo{person}{Michael Isard},
  {et~al\mbox{.}}} \bibinfo{year}{2016}\natexlab{}.
\newblock \showarticletitle{{TensorFlow: A System for Large-Scale Machine
  Learning}}. In \bibinfo{booktitle}{\emph{12th USENIX Symposium on Operating
  Systems Design and Implementation (OSDI 16)}}. \bibinfo{pages}{265--283}.
\newblock


\bibitem[\protect\citeauthoryear{Abu-El-Haija, Kothari, Lee, Natsev, Toderici,
  Varadarajan, and Vijayanarasimhan}{Abu-El-Haija et~al\mbox{.}}{2016}]%
        {YouTube8M}
\bibfield{author}{\bibinfo{person}{Sami Abu-El-Haija}, \bibinfo{person}{Nisarg
  Kothari}, \bibinfo{person}{Joonseok Lee}, \bibinfo{person}{Paul Natsev},
  \bibinfo{person}{George Toderici}, \bibinfo{person}{Balakrishnan
  Varadarajan}, {and} \bibinfo{person}{Sudheendra Vijayanarasimhan}.}
  \bibinfo{year}{2016}\natexlab{}.
\newblock \showarticletitle{{Youtube-8M: A Large-Scale Video Classification
  Benchmark}}.
\newblock \bibinfo{journal}{\emph{arXiv preprint arXiv:1609.08675}}
  (\bibinfo{year}{2016}).
\newblock


\bibitem[\protect\citeauthoryear{Ba, Kiros, and Hinton}{Ba
  et~al\mbox{.}}{2016}]%
        {layer}
\bibfield{author}{\bibinfo{person}{Jimmy~Lei Ba}, \bibinfo{person}{Jamie~Ryan
  Kiros}, {and} \bibinfo{person}{Geoffrey~E Hinton}.}
  \bibinfo{year}{2016}\natexlab{}.
\newblock \showarticletitle{{Layer Normalization}}.
\newblock \bibinfo{journal}{\emph{arXiv preprint arXiv:1607.06450}}
  (\bibinfo{year}{2016}).
\newblock


\bibitem[\protect\citeauthoryear{Deng, Dong, Socher, Li, Li, and Fei-Fei}{Deng
  et~al\mbox{.}}{2009}]%
        {ImageNet}
\bibfield{author}{\bibinfo{person}{Jia Deng}, \bibinfo{person}{Wei Dong},
  \bibinfo{person}{Richard Socher}, \bibinfo{person}{Li-Ji Li},
  \bibinfo{person}{Kai Li}, {and} \bibinfo{person}{Li Fei-Fei}.}
  \bibinfo{year}{2009}\natexlab{}.
\newblock \showarticletitle{{ImageNet: A Large-Scale Hierarchical Image
  Database}}. In \bibinfo{booktitle}{\emph{CVPR09}}.
\newblock


\bibitem[\protect\citeauthoryear{Gehring, Auli, Grangier, Yarats, and
  Dauphin}{Gehring et~al\mbox{.}}{2017}]%
        {posenc}
\bibfield{author}{\bibinfo{person}{Jonas Gehring}, \bibinfo{person}{Michael
  Auli}, \bibinfo{person}{David Grangier}, \bibinfo{person}{Denis Yarats},
  {and} \bibinfo{person}{Yann~N. Dauphin}.} \bibinfo{year}{2017}\natexlab{}.
\newblock \showarticletitle{{Convolutional Sequence to Sequence Learning}}. In
  \bibinfo{booktitle}{\emph{Proceedings of the 34th International Conference on
  Machine Learning-Volume 70}}. \bibinfo{pages}{1243--1252}.
\newblock


\bibitem[\protect\citeauthoryear{Gemmeke, Ellis, Freedman, Jansen, Lawrence,
  Moore, Plakal, and Ritter}{Gemmeke et~al\mbox{.}}{2017}]%
        {AudioSet}
\bibfield{author}{\bibinfo{person}{Jort~F. Gemmeke},
  \bibinfo{person}{Daniel~P.W. Ellis}, \bibinfo{person}{Dylan Freedman},
  \bibinfo{person}{Aren Jansen}, \bibinfo{person}{Wade Lawrence},
  \bibinfo{person}{R.~Channing Moore}, \bibinfo{person}{Manoj Plakal}, {and}
  \bibinfo{person}{Marvin Ritter}.} \bibinfo{year}{2017}\natexlab{}.
\newblock \showarticletitle{{Audio Set: An ontology and human-labeled dataset
  for audio events}}. In \bibinfo{booktitle}{\emph{2017 IEEE International
  Conference on Acoustics, Speech and Signal Processing (ICASSP)}}.
  \bibinfo{pages}{776--780}.
\newblock


\bibitem[\protect\citeauthoryear{He, Zhang, Ren, and Sun}{He
  et~al\mbox{.}}{2016}]%
        {residual}
\bibfield{author}{\bibinfo{person}{Kaiming He}, \bibinfo{person}{Xiangyu
  Zhang}, \bibinfo{person}{Shaoqing Ren}, {and} \bibinfo{person}{Jian Sun}.}
  \bibinfo{year}{2016}\natexlab{}.
\newblock \showarticletitle{{Deep Residual Learning for Image Recognition}}. In
  \bibinfo{booktitle}{\emph{Proceedings of the IEEE conference on computer
  vision and pattern recognition}}. \bibinfo{pages}{770--778}.
\newblock


\bibitem[\protect\citeauthoryear{Hershey, Chaudhuri, Ellis, Gemmeke, Jansen,
  Moore, Plakal, Platt, Saurous, Seybold, et~al\mbox{.}}{Hershey
  et~al\mbox{.}}{2017}]%
        {vggish}
\bibfield{author}{\bibinfo{person}{Shawn Hershey}, \bibinfo{person}{Sourish
  Chaudhuri}, \bibinfo{person}{Daniel~PW Ellis}, \bibinfo{person}{Jort~F.
  Gemmeke}, \bibinfo{person}{Aren Jansen}, \bibinfo{person}{R.~Channing Moore},
  \bibinfo{person}{Manoj Plakal}, \bibinfo{person}{Devin Platt},
  \bibinfo{person}{Rif~A. Saurous}, \bibinfo{person}{Bryan Seybold},
  {et~al\mbox{.}}} \bibinfo{year}{2017}\natexlab{}.
\newblock \showarticletitle{{CNN Architectures for Large-Scale Audio
  Classification}}. In \bibinfo{booktitle}{\emph{2017 IEEE International
  Conference on Acoustics, Speech and Signal processing (ICASSP)}}.
  \bibinfo{pages}{131--135}.
\newblock


\bibitem[\protect\citeauthoryear{Iqbal, Xu, Kong, and Wang}{Iqbal
  et~al\mbox{.}}{2018}]%
        {capsule}
\bibfield{author}{\bibinfo{person}{Turab Iqbal}, \bibinfo{person}{Yong Xu},
  \bibinfo{person}{Qiuqiang Kong}, {and} \bibinfo{person}{Wenwu Wang}.}
  \bibinfo{year}{2018}\natexlab{}.
\newblock \showarticletitle{{Capsule Routing for Sound Event Detection}}. In
  \bibinfo{booktitle}{\emph{2018 26th European Signal Processing Conference
  (EUSIPCO)}}. \bibinfo{pages}{2255--2259}.
\newblock


\bibitem[\protect\citeauthoryear{Kingma and Ba}{Kingma and Ba}{2014}]%
        {adam}
\bibfield{author}{\bibinfo{person}{Diederik~P. Kingma} {and}
  \bibinfo{person}{Jimmy Ba}.} \bibinfo{year}{2014}\natexlab{}.
\newblock \showarticletitle{{Adam: A Method for Stochastic Optimization}}.
\newblock \bibinfo{journal}{\emph{arXiv preprint arXiv:1412.6980}}
  (\bibinfo{year}{2014}).
\newblock


\bibitem[\protect\citeauthoryear{Mesaros, Heittola, Benetos, Foster, Lagrange,
  Virtanen, and Plumbley}{Mesaros et~al\mbox{.}}{2018}]%
        {DCASE2016}
\bibfield{author}{\bibinfo{person}{Annamaria Mesaros}, \bibinfo{person}{Toni
  Heittola}, \bibinfo{person}{Emmanouil Benetos}, \bibinfo{person}{Peter
  Foster}, \bibinfo{person}{Mathieu Lagrange}, \bibinfo{person}{Tuomas
  Virtanen}, {and} \bibinfo{person}{Mark~D. Plumbley}.}
  \bibinfo{year}{2018}\natexlab{}.
\newblock \showarticletitle{{Detection and Classification of Acoustic Scenes
  and Events: Outcome of the DCASE 2016 Challenge}}.
\newblock \bibinfo{journal}{\emph{IEEE/ACM Transactions on Audio, Speech and
  Language Processing (TASLP)}} \bibinfo{volume}{26}, \bibinfo{number}{2}
  (\bibinfo{year}{2018}), \bibinfo{pages}{379--393}.
\newblock


\bibitem[\protect\citeauthoryear{Mesaros, Heittola, Diment, Elizalde, Shah,
  Vincent, Raj, and Virtanen}{Mesaros et~al\mbox{.}}{2017}]%
        {DCASE2017challenge}
\bibfield{author}{\bibinfo{person}{Annamaria Mesaros}, \bibinfo{person}{Toni
  Heittola}, \bibinfo{person}{Aleksandr Diment}, \bibinfo{person}{Benjamin
  Elizalde}, \bibinfo{person}{Ankit Shah}, \bibinfo{person}{Emmanuel Vincent},
  \bibinfo{person}{Bhiksha Raj}, {and} \bibinfo{person}{Tuomas Virtanen}.}
  \bibinfo{year}{2017}\natexlab{}.
\newblock \showarticletitle{{DCASE 2017 challenge setup: Tasks, datasets and
  baseline system}}. In \bibinfo{booktitle}{\emph{DCASE 2017-Workshop on
  Detection and Classification of Acoustic Scenes and Events}}.
\newblock


\bibitem[\protect\citeauthoryear{Mesaros, Heittola, and Virtanen}{Mesaros
  et~al\mbox{.}}{2016}]%
        {metrics}
\bibfield{author}{\bibinfo{person}{Annamaria Mesaros}, \bibinfo{person}{Toni
  Heittola}, {and} \bibinfo{person}{Tuomas Virtanen}.}
  \bibinfo{year}{2016}\natexlab{}.
\newblock \showarticletitle{{Metrics for Polyphonic Sound Event Detection}}.
\newblock \bibinfo{journal}{\emph{Applied Sciences}} \bibinfo{volume}{6},
  \bibinfo{number}{6} (\bibinfo{year}{2016}), \bibinfo{pages}{162}.
\newblock


\bibitem[\protect\citeauthoryear{Parekh, Essid, Ozerov, Duong, P{\'e}rez, and
  Richard}{Parekh et~al\mbox{.}}{2018}]%
        {weakly}
\bibfield{author}{\bibinfo{person}{Sanjeel Parekh}, \bibinfo{person}{Slim
  Essid}, \bibinfo{person}{Alexey Ozerov}, \bibinfo{person}{Ngoc~QK Duong},
  \bibinfo{person}{Patrick P{\'e}rez}, {and} \bibinfo{person}{Ga{\"e}l
  Richard}.} \bibinfo{year}{2018}\natexlab{}.
\newblock \showarticletitle{{Weakly Supervised Representation Learning for
  Unsynchronized Audio-Visual Events}}. In
  \bibinfo{booktitle}{\emph{Proceedings of the IEEE Conference on Computer
  Vision and Pattern Recognition Workshops}}. \bibinfo{pages}{2518--2519}.
\newblock


\bibitem[\protect\citeauthoryear{Plumbley, Kroos, Bello, Richard, Ellis, and
  Mesaros}{Plumbley et~al\mbox{.}}{2018}]%
        {DCASE2018proceedings}
\bibfield{author}{\bibinfo{person}{Mark~D. Plumbley},
  \bibinfo{person}{Christian Kroos}, \bibinfo{person}{Juan~P. Bello},
  \bibinfo{person}{Gaël Richard}, \bibinfo{person}{Daniel~P.W. Ellis}, {and}
  \bibinfo{person}{Annamaria Mesaros}.} \bibinfo{year}{2018}\natexlab{}.
\newblock \bibinfo{booktitle}{\emph{{Proceedings of the Detection and
  Classification of Acoustic Scenes and Events 2018 Workshop (DCASE2018)}}}.
\newblock


\bibitem[\protect\citeauthoryear{Simonyan and Zisserman}{Simonyan and
  Zisserman}{2015}]%
        {VGG16}
\bibfield{author}{\bibinfo{person}{Karen Simonyan} {and}
  \bibinfo{person}{Andrew Zisserman}.} \bibinfo{year}{2015}\natexlab{}.
\newblock \showarticletitle{{Very Deep Convolutional Networks for Large-Scale
  Image Recognition}}. In \bibinfo{booktitle}{\emph{International Conference on
  Learning Representations}}.
\newblock


\bibitem[\protect\citeauthoryear{Srivastava, Hinton, Krizhevsky, Sutskever, and
  Salakhutdinov}{Srivastava et~al\mbox{.}}{2014}]%
        {dropout}
\bibfield{author}{\bibinfo{person}{Nitish Srivastava},
  \bibinfo{person}{Geoffrey Hinton}, \bibinfo{person}{Alex Krizhevsky},
  \bibinfo{person}{Ilya Sutskever}, {and} \bibinfo{person}{Ruslan
  Salakhutdinov}.} \bibinfo{year}{2014}\natexlab{}.
\newblock \showarticletitle{{Dropout: A Simple Way to Prevent Neural Networks
  from Overfitting}}.
\newblock \bibinfo{journal}{\emph{The Journal of Machine Learning Research}}
  \bibinfo{volume}{15}, \bibinfo{number}{1} (\bibinfo{year}{2014}),
  \bibinfo{pages}{1929--1958}.
\newblock


\bibitem[\protect\citeauthoryear{Vaswani, Shazeer, Parmar, Uszkoreit, Jones,
  Gomez, Kaiser, and Polosukhin}{Vaswani et~al\mbox{.}}{2017}]%
        {transformers}
\bibfield{author}{\bibinfo{person}{Ashish Vaswani}, \bibinfo{person}{Noam
  Shazeer}, \bibinfo{person}{Niki Parmar}, \bibinfo{person}{Jakob Uszkoreit},
  \bibinfo{person}{Llion Jones}, \bibinfo{person}{Aidan~N Gomez},
  \bibinfo{person}{{\L}ukasz Kaiser}, {and} \bibinfo{person}{Illia
  Polosukhin}.} \bibinfo{year}{2017}\natexlab{}.
\newblock \showarticletitle{{Attention Is All You Need}}. In
  \bibinfo{booktitle}{\emph{Advances in Neural Information Processing
  Systems}}. \bibinfo{pages}{5998--6008}.
\newblock


\bibitem[\protect\citeauthoryear{Virtanen, Mesaros, Heittola, Diment, Vincent,
  Benetos, and Elizalde}{Virtanen et~al\mbox{.}}{2017}]%
        {DCASE2017proceedings}
\bibfield{author}{\bibinfo{person}{Tuomas Virtanen}, \bibinfo{person}{Annamaria
  Mesaros}, \bibinfo{person}{Toni Heittola}, \bibinfo{person}{Aleksandr
  Diment}, \bibinfo{person}{Emmanuel Vincent}, \bibinfo{person}{Emmanouil
  Benetos}, {and} \bibinfo{person}{Benjamin~Martinez Elizalde}.}
  \bibinfo{year}{2017}\natexlab{}.
\newblock \bibinfo{booktitle}{\emph{{Proceedings of the Detection and
  Classification of Acoustic Scenes and Events 2017 Workshop (DCASE2017)}}}.
\newblock


\bibitem[\protect\citeauthoryear{Xu, Kong, Wang, and Plumbley}{Xu
  et~al\mbox{.}}{2017}]%
        {Surrey}
\bibfield{author}{\bibinfo{person}{Yong Xu}, \bibinfo{person}{Qiuqiang Kong},
  \bibinfo{person}{Wenwu Wang}, {and} \bibinfo{person}{Mark~D Plumbley}.}
  \bibinfo{year}{2017}\natexlab{}.
\newblock \showarticletitle{{Surrey-CVSSP system for DCASE2017 challenge
  task4}}.
\newblock \bibinfo{journal}{\emph{arXiv preprint arXiv:1709.00551}}
  (\bibinfo{year}{2017}).
\newblock


\bibitem[\protect\citeauthoryear{Xu, Kong, Wang, and Plumbley}{Xu
  et~al\mbox{.}}{2018}]%
        {gatedconv}
\bibfield{author}{\bibinfo{person}{Yong Xu}, \bibinfo{person}{Qiuqiang Kong},
  \bibinfo{person}{Wenwu Wang}, {and} \bibinfo{person}{Mark~D Plumbley}.}
  \bibinfo{year}{2018}\natexlab{}.
\newblock \showarticletitle{{Large-scale weakly supervised audio classification
  using gated convolutional neural network}}. In \bibinfo{booktitle}{\emph{2018
  IEEE International Conference on Acoustics, Speech and Signal Processing
  (ICASSP)}}. \bibinfo{pages}{121--125}.
\newblock


\end{thebibliography}
